\documentclass[twocolumn,twoside,preprintnumbers,supercriptaddress,amsmath,amssymb]{revtex4}
 \usepackage{epsfig}
\usepackage{pstcol,times,color,pifont,graphicx,graphics,pstricks,pst-node,pst-coil}
 \usepackage{dcolumn}
 \usepackage{bm}
 \usepackage{fancyhdr}
 \usepackage{pslatex}
 
\begin{document}

 \title{\Large The Casimir effect: some aspects}

 \author{Carlos Farina}
 \affiliation{(1) Universidade Federal do Rio de Janeiro,
 Ilha do Fund\~ao, Caixa Postal 68528, Rio de Janeiro, RJ, 21941-972, Brazil}


 \received{}

 \begin{abstract}
We  start this paper with a historical survey of the Casimir effect,
showing that its origin is related to experiments on colloidal
chemistry. We present two methods of computing Casimir forces,
namely: the global method introduced by Casimir, based on the idea
of zero-point energy of the quantum electromagnetic field, and a
local one, which requires the computation of the energy-momentum
stress tensor of the corresponding field. As explicit examples, we
calculate the (standard) Casimir forces between two parallel and
perfectly conducting plates and discuss the more involved problem of
a scalar field submitted to Robin boundary conditions at two
parallel plates. A few comments are made about recent experiments
that undoubtedly confirm the existence of this effect. Finally, we
briefly discuss a few topics which are either elaborations of the
Casimir effect or topics that are related in some way to this effect
as, for example, the influence of a magnetic field on the Casimir
effect of charged fields, magnetic properties of a confined vacuum
and radiation reaction  forces on non-relativistic moving
boundaries.

 \end{abstract}

 \maketitle
 \thispagestyle{fancy}
\section{Introduction}
\subsection{Some history}

\indent

 The standard Casimir effect was proposed theoretically by
the dutch physicist and humanist Hendrik Brugt Gerhard Casimir
(1909-2000) in 1948 and consists, basically, in the attraction of
two parallel and perfectly conducting plates located in vacuum
\cite{Casimir1948}. As we shall see, this effect has its origin in
colloidal chemistry and is directly related to the dispersive van
der Waals interaction in the retarded regime.

The correct explanation for the non-retarded dispersive van der
Walls interaction between two neutral but polarizable atoms was
possible only after  quantum mechanics was properly established.
Using a perturbative approach, London showed in 1930 \cite{London}
for the first time that the above mentioned interaction is given by
 $ V_{\it Lon}(r)\approx -(3/4)(\hbar\omega_0\alpha^2)/r^6$,
 where $\alpha$ is the static polarizability of the atom,
$\omega_0$ is the dominant transition frequency and $r$
 is the distance between the atoms. In the 40's, various
 experiments with the purpose
of studying equilibrium in colloidal suspensions were made by Verwey
and Overbeek \cite{VerweyOverbeek}. Basically, two types of force
used to be invoked to explain this equilibrium, namely: a repulsive
electrostatic force between layers of charged particles adsorbed by
the colloidal particles and the attractive London-van der Waals
forces.

However, the experiments performed by these authors
 showed that London's interaction  was not correct for large
 distances. Agreement between experimental data and theory was
 possible only if they assumed that the van der Waals interaction
 fell with the distance between two atoms more rapidly than $1/r^6$.
 They even conjectured that the reason for such a different
 behaviour for large distances was due to the retardation effects of the
 electromagnetic interaction (the information of any change or fluctuation
 occurred in one atom should spend a finite time to reach the other
 one). Retardation effects must be taken into account whenever the
 time interval spent by a light signal to travel from one atom to
 the other is of the order (or greater) than atomic characteristic times
 ($r/c\ge 1/\omega_{Man}$, where $\omega_{mn}$ are atomic transition frequencies).
 Though  this conjecture seemed to be very plausible, a rigorous demonstration
 was in order. Further, the precise expression of the van der Waals
 interaction for large distances (retarded regime) should be
 obtained.

Motivated by the disagreement between experiments and theory
described above, Casimir and Polder \cite{CasiPolder48} considered
for the first time, in 1948, the influence of retardation effects on
the van der Waals forces between two atoms as well as on the force
between an atom and a perfectly conducting wall. These authors
obtained their results after lengthy calculations in the context of
perturbative quantum electrodynamics (QED). Since Casimir and
Polder's paper, retarded forces between atoms or molecules and walls
of any kind are usually called {\it Casimir-Polder forces}. They
showed that in the retarded regime the van der Waals interaction
potential between two atoms is given by $ V_{Ret}(r)=-23\hbar
c\alpha_A\alpha_B/(4\pi r^7)$. In contrast to London's result, it
falls as $1/r^7$.
%
 The change in the power law of the dispersive van der Waals force
 when we go from the non-retarded regime to the retarded one
 \linebreak
 ($F_{NR}\sim 1/r^7\; \rightarrow\; F_R\sim 1/r^8$) was measured
 in an experiment with sheets of mica by  D. Tabor and R.H.S. Winterton
 \cite{TaborWinterton1968} only 20 years after Casimir and Polder's
 paper. A change was observed around  $150\hbox{\AA}$, which is the order
 of magnitude of the wavelength of the dominant transition (they worked
 in the range $50\hbox{\AA}-300\hbox{\AA}$, with an accuracy of
 $\pm 4\hbox{\AA}$).
They also showed that the retarded van der Waals interaction
potential  between an atom and a perfectly conducting wall falls as
$1/r^4$, in contrast to the result obtained in the short distance
regime (non-retarded regime), which is proportional to $1/r^3$ (as
can be seen by the image method). Casimir and Polder were very
impressed with the fact that after such a lengthy and involved QED
calculation, the final results were extremely simple. This is very
clear in a conversation with Niels Bohr. In Casimir's own words
\begin{quote}
{\it  In the summer or autumn 1947 (but I am not absolutely certain
that it was not somewhat earlier or later) I mentioned my results to
Niels Bohr, during a walk.\lq\lq That is nice{\rq\rq}, he said,
\lq\lq That is something new.{\rq\rq} I told him that I was puzzled
by the extremely simple form of the expressions for the interaction
at very large distance and he mumbled something about zero-point
energy. That was all,  but it put me on a new track.}
\end{quote}

Following Bohr's suggestion, Casimir re-derived the results obtained
with Polder in a much simpler way, by computing the shift in the
electromagnetic zero-point energy caused by the presence of the
atoms and the walls. He presented his result in the {\it Colloque
sur la th\'eorie de la liaison chimique}, that took place at Paris
in April of 1948:

\begin{quote}
{\it I found that calculating changes of zero-point energy really
leads to the same results as the calculations of Polder and
myself...}
\end{quote}

  A short paper containing this beautiful result was published
  in a French journal only one year later \cite{Casimir1949}.
  Casimir, then, decided to test his method, based on the variation
  of zero-point energy of the electromagnetic field caused by the
  interacting bodies in other examples. He knew that
  the existence of zero-point energy
  of an atomic system (a hot stuff during the years that followed its
  introduction by Planck \cite{Planck1912}) could be inferred by comparing
  energy levels of isotopes. But how to produce isotopes of the
  quantum vacuum? Again, in Casimir's own words we have the answer
  \cite{ProceedingLeipzig98}:

\begin{quote}
{\it if there were two isotopes of empty space you could really easy
confirm the existence of the zero-point energy. Unfortunately, or
perhaps fortunately, there is only one copy of empty space and if
you cannot change the atomic distance then you might change the
shape and that was the idea of the attracting plates.}
\end{quote}

  A month after the {\it Colloque} held at Paris, Casimir
  presented his seminal paper \cite{Casimir1948} on the   attraction between
  two parallel conducting plates which gave   rise to the famous
  effect that since then bears his name:

\begin{quote}
 {\it On 29 May, 1948, \lq I presented my paper
 on the attraction between two perfectly conducting
plates at a meeting of the Royal Netherlands Academy of Arts and
Sciences. It was published in the course of the year...}
\end{quote}

As we shall see explicitly in the next section, Casimir obtained an
attractive force between the plates whose modulus per unit area is
given by
\begin{equation}\label{CasimirForce48}
{F(a)\over L^2} \approx 0,013{1\over \left( a/\mu
m\right)^4}{dyn\over cm^2}\, ,
\end{equation}
where $a$ is the separation between the plates, $L^2$ the area of
each plate (presumably very large, {\it i.e.}, $L\gg a$).

A direct consequence of dispersive van der Waals forces between two
atoms or molecules is that two neutral but polarizable macroscopic
bodies may also interact with each other. However, due to the so
called non-additivity of van der Waals forces, the total interaction
potential between the two bodies is not simply given  by a pairwise
integration, except for the case where the bodies are made of a very
rarefied medium. In principle, the Casimir method provides a way of
obtaining this kind of interaction potential in the retarded regime
(large distances) without the necessity of dealing explicitly with
the non-additivity problem. Retarded van der Waals forces are
usually called Casimir forces. A simple example may be in order.
Consider two semi-infinite slabs made of polarizable material
separated by a distance $a$, as shown in Figure 1.
%
%
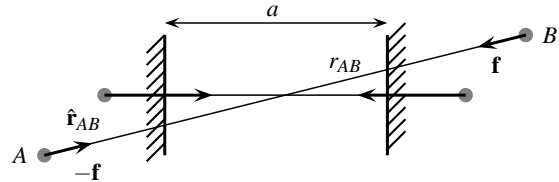
\begin{figure}[!h]
\begin{center}
\newpsobject{showgrid}{psgrid}{subgriddiv=1,griddots=1,gridlabels=6pt}
\begin{pspicture}(0,0)(8,2)
\psset{arrowsize=0.1 2}
\psset{unit=0.8}
%
\pscircle[linewidth=0.9mm,linecolor=gray](1,1){0.12}
\pscircle[linewidth=0.9mm,linecolor=gray](7,1){0.12}
\psline[linewidth=0.2mm](1,1)(7,1)
\psline[linewidth=0.4mm]{->}(1,1)(2.8,1)
\psline[linewidth=0.4mm]{->}(7,1)(5.2,1)
\pscircle[linewidth=0.9mm,linecolor=gray](0,0){0.12}
\pscircle[linewidth=0.9mm,linecolor=gray](8,2){0.12}
\psline[linewidth=0.2mm](0,0)(8,2)
\psline[linewidth=0.4mm]{->}(8,2)(7.2,1.8)
\psline[linewidth=0.4mm]{->}(0,0)(0.8,0.2)
\psline[linewidth=0.4mm](2,2)(2,0)
\psline(2.0,2.0)(1.7,1.7) \psline(2.0,1.8)(1.7,1.5)
\psline(2.0,1.6)(1.7,1.3) \psline(2.0,1.4)(1.7,1.1)
\psline(2.0,1.2)(1.7,0.9) \psline(2.0,1.0)(1.7,0.7)
\psline(2.0,0.8)(1.7,0.5) \psline(2.0,0.6)(1.7,0.3)
\psline(2.0,0.4)(1.7,0.1) \psline(2.0,0.2)(1.7,-0.1)
%
\psline[linewidth=0.4mm](5.7,2)(5.7,0)
\psline(6.0,2.2)(5.7,1.9)
\psline(6.0,2.0)(5.7,1.7) \psline(6.0,1.8)(5.7,1.5)
\psline(6.0,1.6)(5.7,1.3) \psline(6.0,1.4)(5.7,1.1)
\psline(6.0,1.2)(5.7,0.9) \psline(6.0,1.0)(5.7,0.7)
\psline(6.0,0.8)(5.7,0.5) \psline(6.0,0.6)(5.7,0.3)
\psline(6.0,0.4)(5.7,0.1)
%
%
\psline[linewidth=0.2mm]{<->}(2,2.2)(5.7,2.2)
 \rput(3.8,2.4){$a$}
 \rput(-0.4,0){$A$}
 \rput(8.4,2){$B$}
 \rput(0.65,0.6){$\hat{\bf r}_{AB}$}
  \rput(0.7,-0.3){$-{\bf f}$}
 \rput(7.5,1.5){${\bf f}$}
 \rput(5.0,1.5){$r_{AB}$}
\psline[linecolor=white,linewidth=4mm](1.2,2.15)(0.5,2.15)
\psline[linecolor=white,linewidth=5.5mm](7.2,-0.4)(7.2,0.1)
\label{FigInfinteSlabs}
\end{pspicture}
\end{center}
\caption{Forces between molecules of the left slab and molecules of the right slab.}
\end{figure}
%

Suppose the force exerted by a molecule $A$ of the left slab on a
molecule $B$ of the right slab is given by
$$
{\bf f}_{AB} = -\frac{C}{r_{AB}^\gamma}\, {\bf r}_{AB}\, ,
$$
where $C$ and $\gamma$ are positive constants, $r_{AB}$ the distance
between the molecules and ${\hat{\bf r}}_{AB}$ the unit vector
pointing from $A$ to $B$. Hence, by a direct integration it is
straightforward to show that, for the case of dilute media, the
force per unit area between the slabs is attractive and with modulus
given by
\begin{equation}
{F_{\it slabs}\over Area}={C^{\,\prime}\over a^{\gamma-4}}\, ,
\end{equation}
where $C^{\,\prime}$ is a positive constant. Observe that for
$\gamma=8$, which corresponds to the  Casimir and Polder force, we
obtain a force between the slabs per unit area which is proportional
to $1/a^4$. Had we used the Casimir method based on zero-point
energy to compute this force we would have obtained precisely this
kind of dependence. Of course, the numerical coefficients would be
different, since here we made a pairwise integration, neglecting the
non-additivity problem. A detailed discussion on the identification
of the Casimir energy  with the sum of van der Waals interaction for
a dilute dielectric sphere can be found in Milton's book
\cite{MiltonLivro2001} (see also references therein).

In 1956, Lifshitz and collaborators developed a general theory of
van der Waals forces \cite{LifshitzJETP1956}. They derived a
powerful expression for the force at zero temperature as well as at
finite temperature between two semi-infinite dispersive media
characterized by well defined dielectric constants and separated by
a slab of any other dispersive medium. They were able to derive and
predict several results, like the variation of the thickness of thin
superfluid helium films in a remarkable agreement with the
experiments \cite{SabiskyAndersonPRA1973}. The Casimir result for
metallic plates can be reobtained from Lifshitz formula in the
appropriate limit. The Casimir and Polder force can also be inferred
from this formula \cite{MiltonLivro2001} if we consider one of the
media sufficiently dilute such that the force between the slabs may
be obtained by direct integration of a single atom-wall interaction
\cite{DLP}.

The first experimental attempt to verify  the existence of the
Casimir effect for two parallel metallic plates was made by Sparnaay
\cite{SparnaayPhysica1958} only ten years after Casimir's
theoretical prediction. However, due to a very poor accuracy
achieved in this experiment, only compatibility between experimental
data and theory was established. One of the great difficulties was
to maintain a perfect parallelism between the plates. Four decades
have passed, approximately, until new experiments were made directly
with metals. In 1997, using a torsion pendulum Lamoreaux
\cite{LamoreauxPRL1997} inaugurated the new era of experiments
concerning the Casimir effect. Avoiding the parallelism problem, he
measured the Casimir force between a plate and a spherical lens
within the proximity force approximation
\cite{TeoremadaProximidade}. This experiment may be considered a
landmark in the history of the Casimir effect, since it provided the
first reliable experimental confirmation of this effect. One year
later, using an atomic force microscope , Mohideen and Roy
\cite{MohideenRoyPRL1998} measured the Casimir force between a plate
and a sphere with a better accuracy and established an agreement
between experimental data and theoretical predictions of less than a
few percents (depending on the range of distances considered). The
two precise experiments mentioned above have been followed by many
others and an incomplete list of the modern series of experiments
about the Casimir effect can be found in
\cite{KlimchiskayaPRA1999}-\cite{DeccaPRD2003}. For a detailed analysis comparing
theory and experiments  see \cite{DeccaEtAlAnnPhys2005,KlimchitskayaJPA2006}

We finish this subsection emphasizing that Casimir's original
predictions were made for an extremely idealized situation, namely:
two perfectly conducting (flat) plates at zero temperature. Since
the experimental accuracy achieved nowadays is very high, any
attempt to compare theory and experimental data must take into
account more realistic boundary conditions. The most relevant ones
are those that consider the finite conductivity of real metals and
roughness of the surfaces involved. These conditions become more
important as the distance between the two bodies becomes smaller.
Thermal effects must also be considered. However,  in principle,
these effects become dominant compared with the vacuum contribution
for large distances, where the forces are already very small. A
great number of papers have been written on these topics since the
analysis of most recent experiments require the consideration of
real boundary conditions. For finite conductivity effects see Ref.
\cite{LambrechtReynaudEJP2000}; the simultaneous consideration of
roughness and finite conductivity in the proximity for approximation
can be found in Ref. \cite{KlimchitskayaEtAlPRA1999} and beyond PFA
in Ref. \cite{PAMNEtAlPRA2005}
 (see also references cited in the above ones). Concerning the present status of controversies
about  the thermal Casimir force see Ref. \cite{MostepanenkoJPA2006}

\subsection{ The Casimir's approach}

\indent

The  novelty of Casimir's original paper was not the prediction of
an attractive force between neutral objects, once London had already
explained the existence  of a force between neutral but polarizable
atoms, but the method employed by Casimir, which was based on the
zero-point energy ot the electromagnetic field. Proceeding with the
canonical quantization of the electromagnetic field without sources
in the Coulomb gauge we write the hamiltonian operator for the free
radiation field as
\begin{equation}
 \hat H=\sum_{\alpha=1}^2 \sum_{{\bf k}}\hbar\omega_{{\bf k}}
\left[ {\hat a}^\dagger_{{\bf k}\alpha}
 {\hat a}_{{\bf k}\alpha}+{1\over 2}\right]\, ,
\end{equation}
where  ${\hat a}^\dagger_{{\bf k}\alpha}$ and
 ${\hat a}_{{\bf k}\alpha}$ are the creation and annihilation operators of a
 photon with momentum ${\bf k}$ and polarization $\alpha$. The
 energy of the field when it is in the vacuum state, or simply the
vacuum energy, is then given by
\begin{equation}
{\cal E}_0{:=}\langle 0\vert\hat H\vert 0\rangle = \sum_{{\bf
k}}\sum_{\alpha=1}^2 \frac{1}{2}\,
 \hbar\omega_{{\bf k}}\, ,
\end{equation}
which is also referred to as zero-point energy of the
electromagnetic field in free space. Hence, we see that even if we
do not have any real photon in a given mode, this mode will still
contribute to the energy of the field with
$\frac{1}{2}\hbar\omega_{{\bf k}\alpha}$ and total vacuum energy is
then a divergent quantity given by an infinite sum over all possible
modes.

The presence of two parallel and perfectly conducting plates imposes
on the electromagnetic field the following boundary conditions:
\begin{eqnarray}
{\bf E}\times\hat{\bf n}\vert_{plates}&=& {\bf 0}\;\;\;\;\cr
&{\;}&\\
 {\bf B}\cdot\hat{\bf n}\vert_{plates}&=&0\; ,\nonumber
\end{eqnarray}
which modify the possible frequencies of the field modes. The
Casimir energy is, then,  defined as the difference between the
vacuum energy with and without the material plates. However, since
in both situations the vacuum energy is a divergent quantity, we
need to adopt a regularization prescription to give a physical
meaning to such a difference. Therefore, a precise definition for
the Casimir energy is given by
\begin{equation}\label{EnergiaPontoZero}
{\cal E}_{Cas}:=\lim_{s\rightarrow 0}
 \left[\left(\sum_{{\bf k}\alpha} \mbox{${1\over 2}$}\;
 \hbar\omega_{{\bf k}}\right)_I-
\left(\sum_{{\bf k}\alpha} \mbox{${1\over 2}$}\; \hbar\omega_{{\bf
k}}\right)_{II}\right]\; ,
\end{equation}
where subscript $I$ means a regularized sum and that the frequencies
are computed  with the boundary conditions taken into account,
subscript $II$ means a regularized sum but with no boundary
conditions at all and  $s$ stands for the regularizing parameter.
 This definition is well suited for plane geometries like that
analyzed by Casimir in his original work. In more complex
situations, like those involving spherical shells, there are some
subtleties that are beyond the purposes of this introductory article
(the self-energy of a spherical shell depends on its radius while
the self-energy of a pair of plates is independent of the distance
between them).

 Observe that, in the previous definition, we eliminate the regularization
 prescription only after the subtraction is made. Of course, there are many different
regularization methods. A quite simple  but very efficient one is
achieved by introducing a high frequency cut off in the zero-point
energy expression, as we shall see explicitly in the next section.
This procedure can be physically justified if we note that the
metallic plates become transparent in the high frequency limit so
that the high frequency contributions are canceled out from equation
(\ref{EnergiaPontoZero}).

Though the calculation of the Casimir pressure for the case of
 two parallel plates is very simple, its determination may become very involved
 for other geometries, as is the case, for instance, of a perfectly conducting
 spherical shell. After a couple of years of hard work and a
 \lq\lq nightmare in Bessel functions{\rq\rq}, Boyer
 \cite{BoyerPR1968} computed for  the first time the Casimir pressure
 inside a spherical shell. Surprisingly, he found a repulsive
 pressure, contrary to what Casimir had conjectured five years
 before when he proposed a very peculiar model for the stability of
 the electron \cite{CasimirPhysica1953}. Since then, Boyer's result
 has been  confirmed and improved numerically by many authors, as for instance,
 by Davies in 1972 \cite{DaviesJMP1972}, by Balian and Duplantier
 in 1978 \cite{BalianDuplantierAP1978} and also Milton in 1978
 \cite{MiltonRaadSchwingerAP1978}, just to mention some old results.

The Casimir effect is not a peculiarity of the electromagnetic
field. It can be shown that any relativistic field under boundary
conditions caused by material bodies or by a compactification of
space dimensions has its zero-point energy modified. Nowadays, we
denominate by Casimir effect any change in the vacuum energy of a
quantum field due to any external agent, from classical backgrounds
and non-trivial topology to external fields or neighboring bodies.
Detailed reviews of the Casimir effect can be found in
\cite{MiltonLivro2001,PlunienMullerGreiner96,MilonniLivro1994,MostepanenkoTrunov1997,BordagEtAl2001}.
%
%
%
%
\subsection{A local approach}

\indent

In this section we present an alternative way of computing the
Casimir energy density or directly the Casimir pressure which makes
use of a local quantity, namely, the energy-momentum tensor. Recall
that in classical electromagnetism the total force on a distribution
of charges and currents can be computed integrating the Maxwell
stress tensor through an appropriate closed surface containing the
distribution. For simplicity, let us illustrate the method in a
scalar field. The lagrangian density for a free scalar field is
given by
\begin{equation}
{\cal L}\bigl( \phi,\partial_\mu\phi\bigr) =
-\frac{1}{2}\partial_\mu\phi
\partial^\mu\phi - \frac{1}{2} m^2\phi^2
\end{equation}
The field equation and the corresponding Green function are given,
respectively, by
\begin{eqnarray}
(-\partial^2 + m^2)\phi(x)&=&0\; ;\;\\\cr
(\partial^2 - m^2)G(x,x^{\,\prime}) &=& -\delta(x-x^{\,\prime})\, ,
\end{eqnarray}
where, as usual, $G(x,x^{\,\prime}) = i\langle 0\vert
T\Bigl(\phi(x)\phi(x^{\,\prime})\Bigr)\vert 0\rangle$.

Since the above lagrangian density does not depend explicitly on
$x$, Noether's Theorem leads naturally to the following
energy-momentum tensor ($\partial_\mu T^{\mu\nu}=0$)
\begin{equation}
T^{\mu\nu} = \frac{\partial{\cal L}}
 {\partial(\partial_\mu\phi)}\partial^\nu\phi +
 g^{\mu\nu}{\cal L}\, ,
 \end{equation}
which, after symmetrization, can be written in the form
\begin{equation}
T_{\mu\nu} = \frac{1}{2}\Bigl(\partial_\mu\phi\partial_\nu\phi +
\partial_\nu\phi\partial_\mu\phi\Bigr) + g_{\mu\nu}{\cal L}\, .
\end{equation}
For our purposes, it is convenient to write the vacuum expectation
value (VEV) of the energy-momentum tensor in terms of the above
Green function as
\begin{equation}\label{TmunuGreen}
\langle 0\vert T_{\mu\nu}(x)\vert 0\rangle = - \frac{i}{2}
\lim_{x^{\,\prime}\rightarrow x}
\Bigl[(\partial_\mu^\prime\partial_\nu\! +\!
\partial_\nu^\prime\partial_\mu) -
g_{\mu\nu}(\partial_\alpha^\prime\partial^\alpha\!+\! m^2)\Bigr] \!
G(x,x^{\,\prime})\, .
\end{equation}
In this context, the Casimir energy density is defined as
\begin{equation}
\rho_C(x) = \langle 0\vert T_{00}(x)\vert 0\rangle_{BC} - \langle
0\vert T_{00}(x)\vert 0\rangle_{Free}\, ,
\end{equation}
where the subscript $BC$ means that the VEV must be computed
assuming that the field satisfies the appropriate boundary
condition. Analogously, considering for instance the case of two
parallel plates perpendicular to the ${\cal OZ}$ axis (the
generalization for other configurations is straightforward) the
Casimir force per unit area on one plate is given by
\begin{equation}
{\cal F}_C = \langle 0\vert T_{zz}^+\vert 0\rangle -
 \langle 0\vert T_{zz}^-\vert 0\rangle\, ,
\end{equation}
where superscripts $+$ and $-$ mean that we must evaluate $\langle
T_{zz}\rangle$  on both sides of the plate. In other words, the
desired Casimir pressure on the plate is given by the discontinuity
of $\langle T_{zz}\rangle$ at the plate. Using equation
(\ref{TmunuGreen}), $\langle T_{zz}\rangle$ can be computed by
\begin{equation}
 \langle 0\vert T_{zz}\vert 0\rangle = -\frac{i}{2}
\lim_{x^{\,\prime}\rightarrow x}\left(\frac{\partial}{\partial z}
\frac{\partial}{\partial z^{\,\prime}} - \frac{\partial^2}{\partial
z^2}\right) G(x,x^{\,\prime})\, .
\end{equation}

Local methods are richer than global ones, since they provide much
more information about the system. Depending on the problem we are
interested in, they are indeed necessary, as for instance in the
study of radiative properties of an atom inside a cavity. However,
with the purpose of computing Casimir energies in simple situations,
one may choose, for convenience, global methods. Previously, we
presented only the global method introduced by Casimir, based on the
zero-point energy of the quantized field, but there are many others,
namely, the generalized zeta function method \cite{ElizaldeLivroZeta}
and Schwinger's method \cite{Schwinger92,SchwingerGrupoCasimir}, to mention just a
few.

%

\section{Explicit computation of the Casimir force}

\indent

In this section, we show explicitly two ways of computing the
Casimir force per unit area in simple situations where  plane
surfaces are involved. We start with  the global approach introduced
by Casimir which is based on the zero-point energy. Then, we give a
second example where we use a local approach, based on the
energy-momentum tensor. We finish this section by sketching some
results concerning the Casimir effect for massive fields.

\subsection{The electromagnetic Casimir effect between two  parallel
plates}

\indent

As our first example, let us consider the standard (QED) Casimir
effect  where the quantized electromagnetic field is constrained by
two perfectly parallel conducting plates separated by a distance
$a$. For convenience, let us suppose that one plate is located at
 $z=0$, while the other is located at $z=a$. The quantum electromagnetic
 potential between the metallic plates in the Coulomb
gauge ($\nabla\cdot{\bf A}=0$) which satisfies the appropriate BC is
given by \cite{Barton70}

\begin{eqnarray}
{\bf{A}}(\mbox{{\mathversion{bold}${\rho}$}} ,z,t)&=&
\frac{L^2}{(2\pi)^2}{\sum_{n=0}^{\infty\;\;\prime}} \int
 d^2\mbox{{\mathversion{bold}${\kappa}$}}
 \left(\frac{2\pi\hbar}{ckaL^2}\right)^{1/2}\times\cr\cr
&\times& \Biggl\{ a^{(1)}
 (\mbox{{\mathversion{bold}${\kappa}$}},n)
 (\hat{\mbox{{\mathversion{bold}${\kappa}$}}}\times\hat{\bf z})
\sin\left(\frac{n\pi z}{a}\right) + \cr\cr
 &+& a^{(2)}\!
 (\mbox{{\mathversion{bold}${\kappa}$}},n)\!\!
 \left[i\frac{n\pi}{ka}
 \mbox{{\mathversion{bold}${\kappa}$}}
 \sin\left(\frac{n\pi z}{a}\right)
 - \hat{\bf z}\frac{\kappa}{k} \cos\left(\frac{n\pi z}{a}\right)\right]
 \Biggr\}\times\cr\cr
 &\times&
 e^{i(\mbox{{\mathversion{bold}${\kappa}$}}\cdot
 \mbox{{\mathversion{bold}${\rho}$}}
 - \omega\, t)} \; +\; h.c.\; ,
\end{eqnarray}
where $\omega(\kappa,n) = ck = c\,\left[ \kappa^2 +
n^2\pi^2/a^2\right]^{1/2}$, with $n$ being a non-negative integer and the prime in
$\Sigma^{\,\prime}$ means that for $n=0$  an extra $1/2$ factor must be included
 in the normalization of the field modes. The  non-regularized Casimir energy then  reads
\begin{eqnarray}
{\cal E}_c^{nr}(a)&=&{\hbar c\over 2}\int L^2
 {d^2{\mbox{{\mathversion{bold}${\kappa}$}}}
 \over
(2\pi)^2}\left[ \kappa + 2 \sum_{n=1}^\infty\left({\kappa}^2+ {n^2
\pi^2\over a^2}\right)^{1/2}\right]\nonumber\cr\cr &{\;}&\;\;\;\;\;-
\;\;{\hbar c\over 2}\int L^2
{d^2{\mbox{{\mathversion{bold}${\kappa}$}}}\over (2\pi)^2}
\int_{-\infty}^{+\infty} {a dk_z\over 2\pi}\,
2\sqrt{{\kappa}^2+k^2_z}\, .
\end{eqnarray}
Making the variable transformation $\kappa^2 + (n\pi/a)^2 =:
\lambda$ and introducing exponential cutoffs we get a regularized
expression (in 1948 Casimir used a generic cutoff function),
\begin{eqnarray}
{\cal E}^r(a,\varepsilon) =
\frac{L^2}{2\pi}
&\Biggl[&
\frac{1}{2}\int_0^\infty
 e^{-\varepsilon\kappa}\kappa^2\, d\kappa +
\sum_{n=1}^\infty\int_{\frac{n\pi}{a}}^\infty
 e^{-\varepsilon\lambda}\lambda^2\, d\lambda \ - \cr\cr
 &-&
 \int_0^\infty dn
\!\!\int_{\frac{n\pi}{a}}^\infty
 e^{-\varepsilon\lambda}\lambda^2 d\lambda\Biggr]\cr\cr\cr
=
 \frac{L^2}{2\pi}
 &\Biggl[&
 \frac{1}{\varepsilon^3} +
\sum_{n=1}^\infty \frac{\partial^2}{\partial \varepsilon^2}
\left(\frac{e^{-\varepsilon n\pi/a}}{\varepsilon}\right) -\cr\cr
&-&
\int_0^\infty dn\frac{\partial^2}{\partial\varepsilon^2}
\int_{\frac{n\pi}{a}}^\infty e^{-\varepsilon\lambda} d\lambda
\,\Biggr]\, .
\end{eqnarray}
Using that
$$
\frac{\partial^2}{\partial \varepsilon^2}\left[\frac{1}{\varepsilon}
\sum_{n=1}^\infty e^{-\varepsilon n\pi/a}\right] =
\frac{\partial^2}{\partial \varepsilon^2}\left[\frac{1}{\varepsilon}
\frac{1}{e^{\varepsilon\pi/a} - 1}\right]\, ,
$$
as well as the definition of Bernoulli's numbers,
$$
\frac{1}{e^t - 1} = \sum_{n=0}^\infty B_n\,\frac{t^{n-1}}{n!}\, ,
$$
we obtain
\begin{eqnarray}
{\cal E}(a)\!\!\!&=&\!\!\!
\frac{L^2}{2\pi}\left[\frac{1}{\varepsilon^3} +
\frac{\partial^2}{\partial \varepsilon^2}
\left\{\frac{1}{\varepsilon}\sum_{n=0}^\infty \frac{B_n}{n!}
\left(\frac{\varepsilon\pi}{a}\right)^{n-1}\right\} -
\frac{6a}{\pi\varepsilon^4}\right]\cr\cr
 \!\!\!&=&\!\!\!
\frac{L^2}{2\pi}\Biggl[ 6(B_0 \!\! - \!\!
1)\frac{a}{\pi}\frac{1}{\varepsilon^4} +
(1+2B_1)\frac{1}{\varepsilon^3} +
\frac{B_4}{12}\left(\frac{\pi}{a}\right)^3\cr\cr
 \!\!\!&+&\!\!\!
\sum_{n=5}^\infty \frac{B_n}{n!}\left(\frac{\pi}{a}\right)^{n-1}
(n-2)(n-3)\varepsilon^{n-4}\Biggr]\, .\nonumber
\end{eqnarray}
Using the well known values $B_0=1,\;
B_1=-\mbox{\large$\frac{1}{2}$}$ and
 $B_4=-\mbox{\large$\frac{1}{30}$}$,
and taking $\varepsilon\rightarrow 0^+$, we obtain
$$
\frac{{\cal E}(a)}{L^2} = -\hbar c\,\frac{\pi^2}{24\times 30}\cdot\frac{1}{a^3}
$$
As a consequence, the force per unit area acting on the plate at
$z=a$ is given by
$$
{F(a)\over L^2} = -{1\over L^2}\frac{\partial {\cal
E}_c(a)}{\partial a} = -{\pi^2\hbar c\over 240 a^4} \approx
-0,013{1\over \left( a/\mu m\right)^4}{dyn\over cm^2}\, ,
$$
where in the last step we substituted the numerical values of
$\hbar$ and $c$ in order to give an idea of the strength of the
Casimir pressure. Observe that the Casimir force between the
(conducting) plates is always attractive. For plates with $1$cm$^2$
of area separated by $1\mu$m the modulus of this attractive force is
$0,013$dyn. For this same separation, we have
 $P_{Cas}\approx 10^{-8}\, P_{atm}$, where $ P_{atm}$ is the
 atmospheric pressure at sea level. Hence, for the idealized
 situation of two perfectly conducting plates and assuming  $L^2=1\; cm^2$,
the modulus of the Casimir force would be $\approx 10^{-7}N$ for
typical separations used in experiments. However, due to the finite
conductivity of real metals, the Casimir forces measured in
experiments are smaller than these values.

\subsection{The Casimir effect for a scalar field with Robin BC}

\indent

In order to illustrate the local method based on the energy-momentum
tensor, we shall discuss the Casimir effect of a massless scalar
field submitted to Robin BC at two parallel plates, which are
defined as
\begin{equation}
\phi\vert_{bound.} = \beta\,\frac{\partial\phi}{\partial
n}\vert_{bound.}\, ,
\end{equation}
where, by assumption, $\beta$ is a non-negative parameter. However,
before computing the desired Casimir pressure, a few comments about
Robin BC are in order.

First, we note that Robin BC interpolate continuously Dirichlet and
Neumann ones. For $\beta\rightarrow 0$ we reobtain Dirichlet BC
while for $\beta\rightarrow\infty$ we reobtain Neumann BC. Robin BC
already appear in classical electromagnetism, classical mechanics,
wave, heat and Schr\"odinger equations \cite{BondurantFulling} and
even in the study of interpolating partition functions
\cite{AsoreyRosaLeo}. A nice  realization of these conditions in the
context of classical mechanics can be obtained if we study a
vibrating string with its extremes attached to elastic supports
\cite{Zhou,Mintz1}. Robin BC can also be used in a phenomenological
model for a penetrable surfaces \cite{MostepanenkoTrunov85}. In
fact, in the context of the plasma model, it can be shown that for
frequencies much smaller than the plasma frequency
$(\omega\ll\omega_P)$ the parameter $\beta$ plays the role of the
plasma wavelength \cite{Mintz2}. In the context of QFT this kind of
BC appeared more than two decades ago
\cite{DeutschCandelas79,KennedyCritchleyDowker80}. Recently, they
have been discussed in a variety of contexts, as in the AdS/CFT
correspondence \cite{MinceRivelles2000}, in the discussion of upper
bounds for the ratio entropy/energy in confined systems
\cite{Solodukihn2001}, in the static Casimir effect
\cite{RomeoSaharian2002}, in the heat kernel expansion
\cite{BordagFalomirSantangeloVassilevich2002,Fulling2003,Dowker2005}
and in the one-loop renormalization of the  $\lambda\phi^4$ theory
\cite{AlbuquerqueCavalcanti2002,Albuquerque2005}.
 As we will see, Robin BC can give rise to restoring
 forces in the static Casimir effect \cite{RomeoSaharian2002}.

 Consider a massless scalar field submitted to Robin BC on two parallel plates:
\begin{equation}
\phi\vert_{z=0} = \beta_1\,\frac{\partial\phi}{\partial
z}\vert_{z=0} \;\; ;\;\;\; \phi\vert_{z=a} =
-\beta_2\,\frac{\partial\phi}{\partial z}\vert_{z=a}\; ,
\end{equation}
where we assume $\beta_1(\beta_2)\geq 0$. For convenience, we write
\begin{equation}
G(x,x')=\int \frac{d^3k_{\alpha}}{(2\pi)^3}e^{\imath
k_{\alpha}(x-x')^{\alpha}}g(z,z^{\,\prime};\textbf{k}_{\perp},\omega),
\end{equation}
with $\alpha =0,1,2$, $g_{\mu \nu}=diag(-1,+1,+1,+1)$ and the
reduced Green function $g(z,z^{\,\prime};\textbf{k}_{\perp},\omega)$
satisfies
\begin{equation}
 \left (\frac{\partial^2}{\partial z^2}+\lambda^2
\right )g(z,z^{\,\prime};\textbf{k}_{\perp},\omega)=-\delta
(z-z^{\,\prime})\, ,
\end{equation}
where $\lambda^2 =\omega^2-\textbf{k}^2_{\perp}$ and
$g(z,z^{\,\prime};\textbf{k}_{\perp},\omega)$ is submitted to the
following boundary conditions (for simplicity, we shall not write
$\textbf{k}_{\perp},\omega$ in the argument of $g$):
$$
g(0,z^{\,\prime})=\beta_1\frac{\partial }{\partial
z}g(0,z^{\,\prime})\;\;\; ; \hspace{1.0cm}
g(a,z^{\,\prime})=-\beta_2\frac{\partial }
 {\partial z}g(a,z^{\,\prime})
$$
It is not difficult to see that $g(z,z^{\,\prime})$ can be written
as
\begin{equation}\nonumber
g(z,z^{\,\prime}) \!=\!\!\left \{ \;\;\begin{array}{ll}
\!\!\!\!\!A(z^{\,\prime})\left ( \sin \lambda z +\beta_1\lambda \cos
\lambda z\right ),& \!\!\!\!z<z^{\,\prime}\cr\cr
\!\!\!\!\!B(z^{\,\prime})\left ( \sin \lambda (z\!-a\!)\!
-\!\beta_2\lambda \cos \lambda (z-a)\right )\!,&
\!\!\!z>z^{\,\prime}
\end{array}
\right.
\end{equation}
For points inside the plates, $0<z,z^{\,\prime}<a$, the final
expression for the reduced Green function is given by
\begin{equation}
g^{RR}(z,z^{\,\prime}) = -\frac{\left (\frac{1}{\gamma_1}e^{\imath
\lambda z_<}-e^{-\imath \lambda z_<} \right )\left
(\frac{1}{\gamma_2}e^{\imath \lambda (z_>-a)}-e^{\imath \lambda
(z_>-a)}\right )}{2\imath \lambda (\frac{1}{\gamma_1}e^{\imath
\lambda a}-\frac{1}{\gamma_2}e^{-\imath \lambda a})}\, ,
\end{equation}
where $\gamma_i=\mbox{\large$\frac{1+ i\, \beta_i \lambda}
 {1- i\,\beta_i\lambda}$}$ ($i=1,2$). Outside the plates,
 with $z,z^{\,\prime}>a$, we have:
\begin{equation}
g^{RR}(z,z^{\,\prime})=\frac{e^{\imath \lambda (z_<-a) }}{2\imath
\lambda}\left (\frac{1}{\gamma_2}e^{\imath \lambda
(z_<-a)}-e^{-\imath \lambda (z_<-a)}\right ).
\end{equation}

Defining $t^{\mu\nu}$ such that
\begin{equation}
\langle T^{\mu\nu}(x)\rangle=\int
\frac{d^2{\bf k}_{\perp}}{(2\pi)}\frac{d\omega}{2\pi}\langle t^{\mu\nu}(x)\rangle\, ,
\end{equation}
we have
\begin{equation}
\langle t^{33}(z)\rangle = \frac{1}{2i}\lim_{z^{\,\prime}\rightarrow z}
\left(\frac{\partial}{\partial z}\frac{\partial}{\partial z^{\,\prime}} + \lambda^2\right)
g(z,z^{\,\prime})\, .
\end{equation}
The Casimir force per unit area on the plate at $x=a$ is given by
 the discontinuity in $\langle t^{33}\rangle$:
$$
 {\cal F}=\int
\frac{d^2{\bf k}_{\perp}}{(2\pi)}\frac{d\omega}{2\pi}
 \Bigl[\langle t^{33}\rangle |_{z=a_-} -\langle t^{33}\rangle
 |_{z=a_+}\Bigr]\, .
$$
After a straightforward calculation, it can be shown that
\begin{equation}
{\cal F}(\beta_1,\!\beta_2;\! a) \! = \! -\frac{1}{32\pi^2a^4}
\!\!\!\int_0^{\infty}\!\!\!\!\!  \frac{d\xi\; \xi^3} {\left(
\frac{1+\beta_1\xi/2a}{1-\beta_1\xi/2a} \right)
\!\!\left(\frac{1+\beta_2\xi/2a}{1-\beta_2\xi/2a} \right )e^\xi \!
-1\! }\, .
\end{equation}
Depending on the values of parameters $\beta_1$ and $\beta_2$, restoring Casimir forces may
arise, as shown in Figure 2 by the dotted line and the thin solid line.

\vskip 0.5cm

\begin{figure}[!h]
\newpsobject{showgrid}{psgrid}{subgriddiv=1,griddots=10,gridlabels=6pt}
\begin{pspicture}(-1,-2.5)(18,3.2)

\psset{arrowsize=0.2 2}
 \psset{unit=0.40}


\psline{->}(-0.5,0)(17.5,0)

\psline{->}(0,-6.5)(0,9)

\rput(-1.2,7.0){\Large$p a^4$}

\rput(17,-0.8){\Large$a$}

\rput(-1,-0.8){\Large${\cal O}$}

\psline [linewidth=.15,linestyle=solid] (0,-6)(15,-6)

\psline [linewidth=.15,linestyle=solid] (0,5.25)(15,5.25)

\pscurve [linewidth=.15,linestyle=dotted] (0., 5.2500)
(.1250000000, 4.9550) (.2500000000, 4.6767) (.3750000000, 4.4138)
(.5000000000, 4.1648) (.6250000000, 3.9285) (.7500000000, 3.7040)
(.8750000000, 3.4903) (1.000000000, 3.2866) (1.125000000, 3.0923)
(1.250000000, 2.9066) (1.375000000, 2.7289) (1.500000000, 2.5586)
(1.625000000, 2.3954) (1.750000000, 2.2388) (1.875000000, 2.0881)
(2.000000000, 1.9433) (2.125000000, 1.8040) (2.250000000, 1.6698)
(2.375000000, 1.5403) (2.500000000, 1.4153) (2.625000000, 1.2948)
(2.750000000, 1.1783) (2.875000000, 1.0657) (3.000000000, .95670)
(3.125000000, .85122) (3.250000000, .74903) (3.375000000, .65001)
(3.500000000, .55399) (3.625000000, .46083) (3.750000000, .37039)
(3.875000000, .28257) (4.000000000, .19724) (4.125000000, .11430)
(4.250000000, 0.033635) (4.375000000, -0.044841) (4.500000000,
-.12122) (4.625000000, -.19559) (4.750000000, -.26803)
(4.875000000, -.33862) (5.000000000, -.40742) (5.125000000,
-.47452) (5.250000000, -.53997) (5.375000000, -.60383)
(5.500000000, -.66617) (5.625000000, -.72703) (5.750000000,
-.78647) (5.875000000, -.84455) (6.000000000, -.90135)
(6.125000000, -.95679) (6.250000000, -1.0111) (6.375000000,
-1.0641) (6.500000000, -1.1160) (6.625000000, -1.1668)
(6.750000000, -1.2165) (6.875000000, -1.2652) (7.000000000,
-1.3129) (7.125000000, -1.3596) (7.250000000, -1.4054)
(7.375000000, -1.4503) (7.500000000, -1.4942) (7.625000000,
-1.5373) (7.750000000, -1.5796) (7.875000000, -1.6210)
(8.000000000, -1.6616) (8.125000000, -1.7015) (8.250000000,
-1.7407) (8.375000000, -1.7791) (8.500000000, -1.8168)
(8.625000000, -1.8539) (8.750000000, -1.8902) (8.875000000,
-1.9259) (9.000000000, -1.9611) (9.125000000, -1.9956)
(9.250000000, -2.0294) (9.375000000, -2.0628) (9.500000000,
-2.0955) (9.625000000, -2.1278) (9.750000000, -2.1594)
(9.875000000, -2.1905) (10.00000000, -2.2211) (10.12500000,
-2.2513) (10.25000000, -2.2810) (10.37500000, -2.3100)
(10.50000000, -2.3387) (10.62500000, -2.3670) (10.75000000,
-2.3948) (10.87500000, -2.4222) (11.00000000, -2.4492)
(11.12500000, -2.4757) (11.25000000, -2.5019) (11.37500000,
-2.5276) (11.50000000, -2.5530) (11.62500000, -2.5780)
(11.75000000, -2.6026) (11.87500000, -2.6269) (12.00000000,
-2.6509) (12.12500000, -2.6745) (12.25000000, -2.6978)
(12.37500000, -2.7207) (12.50000000, -2.7433) (12.62500000,
-2.7656) (12.75000000, -2.7876) (12.87500000, -2.8093)
(13.00000000, -2.8307) (13.12500000, -2.8517) (13.25000000,
-2.8725) (13.37500000, -2.8931) (13.50000000, -2.9134)
(13.62500000, -2.9335) (13.75000000, -2.9531) (13.87500000,
-2.9726) (14.00000000, -2.9918) (14.12500000, -3.0109)
(14.25000000, -3.0296) (14.37500000, -3.0481) (14.50000000,
-3.0664) (14.62500000, -3.0845) (14.75000000, -3.1023)
(14.87500000, -3.1198) (15.00000000, -3.1372)

\pscurve [linewidth=.04,linestyle=solid]  (0., -6.0000)
(.1250000000, -2.1470) (.2500000000, -.58810) (.3750000000,
.27043) (.5000000000, .78780) (.6250000000, 1.1107) (.7500000000,
1.3124) (.8750000000, 1.4341) (1.000000000, 1.5008) (1.125000000,
1.5286) (1.250000000, 1.5284) (1.375000000, 1.5075) (1.500000000,
1.4714) (1.625000000, 1.4239) (1.750000000, 1.3680) (1.875000000,
1.3056) (2.000000000, 1.2388) (2.125000000, 1.1683) (2.250000000,
1.0956) (2.375000000, 1.0212) (2.500000000, .94581) (2.625000000,
.86995) (2.750000000, .79385) (2.875000000, .71793) (3.000000000,
.64238) (3.125000000, .56740) (3.250000000, .49316) (3.375000000,
.41973) (3.500000000, .34726) (3.625000000, .27578) (3.750000000,
.20535) (3.875000000, .13603) (4.000000000, 0.067803)
(4.125000000, 0.00072286) (4.250000000, -0.065211) (4.375000000,
-.13000) (4.500000000, -.19364) (4.625000000, -.25613)
(4.750000000, -.31751) (4.875000000, -.37777) (5.000000000,
-.43690) (5.125000000, -.49496) (5.250000000, -.55195)
(5.375000000, -.60789) (5.500000000, -.66280) (5.625000000,
-.71668) (5.750000000, -.76959) (5.875000000, -.82151)
(6.000000000, -.87249) (6.125000000, -.92250) (6.250000000,
-.97173) (6.375000000, -1.0200) (6.500000000, -1.0673)
(6.625000000, -1.1138) (6.750000000, -1.1596) (6.875000000,
-1.2045) (7.000000000, -1.2485) (7.125000000, -1.2919)
(7.250000000, -1.3343) (7.375000000, -1.3762) (7.500000000,
-1.4173) (7.625000000, -1.4577) (7.750000000, -1.4974)
(7.875000000, -1.5365) (8.000000000, -1.5748) (8.125000000,
-1.6126) (8.250000000, -1.6498) (8.375000000, -1.6862)
(8.500000000, -1.7222) (8.625000000, -1.7575) (8.750000000,
-1.7924) (8.875000000, -1.8266) (9.000000000, -1.8602)
(9.125000000, -1.8933) (9.250000000, -1.9259) (9.375000000,
-1.9581) (9.500000000, -1.9897) (9.625000000, -2.0209)
(9.750000000, -2.0516) (9.875000000, -2.0816) (10.00000000,
-2.1114) (10.12500000, -2.1406) (10.25000000, -2.1696)
(10.37500000, -2.1980) (10.50000000, -2.2261) (10.62500000,
-2.2536) (10.75000000, -2.2809) (10.87500000, -2.3076)
(11.00000000, -2.3341) (11.12500000, -2.3602) (11.25000000,
-2.3859) (11.37500000, -2.4112) (11.50000000, -2.4362)
(11.62500000, -2.4608) (11.75000000, -2.4851) (11.87500000,
-2.5091) (12.00000000, -2.5328) (12.12500000, -2.5562)
(12.25000000, -2.5792) (12.37500000, -2.6019) (12.50000000,
-2.6243) (12.62500000, -2.6464) (12.75000000, -2.6682)
(12.87500000, -2.6898) (13.00000000, -2.7112) (13.12500000,
-2.7321) (13.25000000, -2.7529) (13.37500000, -2.7734)
(13.50000000, -2.7936) (13.62500000, -2.8136) (13.75000000,
-2.8333) (13.87500000, -2.8527) (14.00000000, -2.8721)
(14.12500000, -2.8910) (14.25000000, -2.9099) (14.37500000,
-2.9284) (14.50000000, -2.9467) (14.62500000, -2.9649)
(14.75000000, -2.9828) (14.87500000, -3.0004) (15.00000000,
-3.0180)

\rput(6,6.0){($\beta_1=0\; ,\beta_2\rightarrow\infty$)}
\rput(8,-5){($\beta_1=\beta_2=0\;\mbox{or}\;
\beta_1=\beta_2\rightarrow\infty$)}
\rput(6,3){($\beta_1=0\; ,\;\beta_2=1$)}

\label{FigRobin}
\end{pspicture}
\caption{Casimir pressure, conveniently multiplied by $a^4$, as a function of $a$ for various values of
parameters $\beta_1$ and $\beta_2$.}
\end{figure}
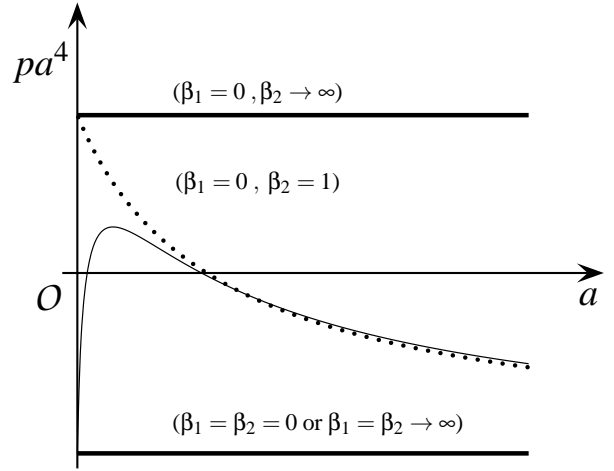

The particular cases of Dirichlet-Dirichlet, Neumann-Neumann and
Dirichlet-Neumann BC  can be reobtained if we take, respectively,
$\beta_1=\beta_2=0$, $\beta_1=\beta_2\rightarrow\infty$ and
$\beta_1=0\, ;\;\beta_2\rightarrow\infty$. For the first two cases, we obtain
\begin{equation}
{\cal F}^{DD}(a)  = {\cal F}^{NN}(a) = -\frac{1}{32\pi^2a^4}
\int_0^{\infty} d\xi\; \frac{ \xi^3} {e^\xi \! -1\! }\, .
\end{equation}
Using the integral representation
 $ \int_0^{\infty} d\xi\;
  \mbox{\large$\frac{\xi^{s-1}}{e^\xi  -1}$} = \zeta_R(s)\Gamma(s)$,
where $\zeta_R$ is the Riemann zeta function, we get half the
electromagnetic result, namely,
\begin{equation}\label{FporAreaMassaNulaEscalar}
{\cal F}^{DD}(a)  = {\cal F}^{NN}(a) = -\frac{\pi^2\hbar c}{480}\frac{1}{a^4}\, .
\end{equation}

For the case of mixed BC, we get
\begin{equation}
{\cal F}^{DN}(a) = + \frac{1}{32\pi^2a^4}
\int_0^{\infty} d\xi\; \frac{ \xi^3}{e^\xi + 1}\; ,
\end{equation}
Using in the previous equation the integral representation
$
 \int_0^{\infty} d\xi\;
 \mbox{\large$\frac{ \xi^{s-1}}{e^\xi + 1}$} =
(1-2^{1-s})\Gamma(s)\zeta_R(s)
$
we get a repulsive pressure (equal to half of Boyer's result
\cite{Boyer74} obtained for the electromagnetic field constrained by
a perfectly conducting plate parallel to an infinitely permeable
one),
\begin{equation}
{\cal F}^{DN}(a)  =  \frac{7}{8}\times\frac{\pi^2\hbar c}{480}\frac{1}{a^4}\, .
\end{equation}

\subsection{The Casimir effect for massive particles}

\indent

In this subsection, we  sketch briefly some results concerning massive fields, just to get some feeling
about what kind of influence the mass of a field may have in the Casimir effect. Firstly, let us consider a massive
scalar field submitted to Dirichlet BC in two parallel plates, as before. In this case, the allowed
frequencies for the field modes are given by
 $\omega_{\bf k}=c\left[\kappa^2+{n^2\pi^2\over a^2}+{m^2}\right]^{1/2}$, which lead, after we use
 the Casimir method explained previously, to the following result
 for the Casimir energy per unit area (see, for instance, Ref. \cite{PlunienMullerGreiner96})
\begin{equation}
{1\over L^2}\;{\cal E}_c(a,m)=- {m^2\over
8\pi^2 a}\;\sum_{n=1}^\infty \frac{1}{n^2} \;
K_2(2amn)\; ,
\end{equation}
where $K_\nu$ is a modified Bessel function, $m$ is the mass of the
field and $a$ is the distance between the plates, as usual. The
limit of small mass, $am\ll 1$, is easily obtained and yields
\begin{equation}
{1\over L^2}\;{\cal E}_c(a,m) \approx - {\pi^2\over 1440\,
a^3}+{m^2\over 96\, a}\; .
\end{equation}
As expected, the zero mass limit coincides with our previous result
(\ref{FporAreaMassaNulaEscalar}) (after  the force per unit area is
computed). Observing the sign of the first correction on the right
hand side of last equation we conclude that for small masses the
Casimir effect is weakened.

On the other hand, in the limit of large mass, $am\gg 1$, it can be
shown that
\begin{equation}
{1\over L^2}\;{\cal E}_c(a,m)\approx
-{m^2\over 16\pi^2\, a}\; \left({\pi\over ma}\right)^{1/2}\, e^{-2ma}
\end{equation}
Note that the Casimir effect disappears for $m\rightarrow\infty$,
since in this limit there are no quantum fluctuations for the field
anymore. An exponential decay with $ma$ is related to the plane
geometry. Other geometries may give rise to power law decays when
$ma\rightarrow\infty$. However, in some cases,
 the behaviour of the Casimir force with $ma$ may be quite unexpected.
 The Casimir force may increase with $ma$ before
 it decreases monotonically to zero as $ma\rightarrow\infty$ (this happens,
 for instance, when Robin BC are imposed on
 a massive scalar field at two parallel plates \cite{ThiagoRobin}).

In the case of massive fermionic fields, an analogous behaviour is
found. However, some care must be taken when computing the Casimir
energy density for fermions, regarding what kind of BC can be chosen
for this field. The point is that Dirac equation is a first order
equation, so that if we want non-trivial solutions, we can not
impose that  the field satisfies Dirichlet BC at two parallel
plates, for instance. The most appropriate BC for fermions is
borrowed from the so called MIT bag model for hadrons
\cite{BagModelChodosPRD1974}, which basically states that there is
no flux of fermions through the boundary (the normal component of
the fermionic current must vanish at the boundary). The Casimir
energy per unit area for a massive fermionic field submitted to MIT
BC at two parallel plates was first computed by Mamayev and Trunov
\cite{MamayevTrunov79}
 (the  massless fermionic field was first computed  by Johnson in 1975
\cite{Johnson75})
\begin{equation}
{\frac{1}{L^2}}{\cal E}_c^f( a,m) = -\frac{1}{\pi^2
a^3}\!\int_{ma}^\infty \! d\xi\, \xi\sqrt{\xi^2 - m^2a^2}
\log\left[1+\frac{\xi -ma} {\xi + ma}e^{-2\xi}\right].\nonumber
\end{equation}
The small and large mass limits are given, respectively, by
\begin{eqnarray}
\frac{1}{L^2}\,{\cal E}_c^f(a,m) &\approx&
-\frac{7\pi^2}{2880 a^3} + \frac{m}{24a^2}\; ;\cr\cr
\frac{1}{L^2}\,{\cal E}_c^f(a,m) &\approx&
-\frac{3(ma)^{1/2}}{2^5\pi^{3/2} a^3} e^{-2ma}\; .
\end{eqnarray}
Since the first correction to the zero mass result has an opposite
sign, also for a fermionic field small masses diminish the Casimir
effect. In the large mass limit, $ma\rightarrow\infty$, we have a
behaviour analogous to that of the scalar field, namely, an
exponential decay with $ma$ (again this happens due to the plane
geometry). We finish this section with an important observation:
even a particle as light as the electron has a completely negligible
Casimir effect.

%
%
\section{Miscellany}

\indent

 In this section we shall  briefly present a couple of topics which
 are in some way connected to the Casimir effect and that have been
 considered by our research group in the last years. For obvious reasons, we
 will not be able to touch all the topics we have been interested in,
 so that we had to choose only a few of them. We first discuss how
 the Casimir effect of a charged field can be influenced by an
 external magnetic field. Then, we show how the constitutive
 equations associated to the Dirac quantum vacuum can be affected by the
 presence of material plates. Finally, we consider the so called
 dynamical Casimir effect.

\subsection{Casimir effect under an external magnetic field}

\indent

The Casimir effect which is observed experimentally is that
associated to the photon field, which is a massless field. As we
mentioned previously, even the electron field already exhibits a
completely unmeasurable effect. With the purpose (and hope) of
enhancing the Casimir effect of electrons and positrons we
considered the influence of an external electromagnetic field on
their Casimir effect. Since in this case we have charged fields, we
wondered if the virtual electron-positron pairs which are
continuously created and destroyed  from the Dirac vacuum would
respond in such a way that the corresponding Casimir energy would be
greatly amplified. This problem was considered for the first time in
1998 \cite{Leipzig98} (see also Ref. \cite{Maldito}).
 The influence of an external magnetic field on the Casimir effect of
 a charged scalar field was considered in Ref. \cite{Guida99} and recently, the influence
  of a magnetic field on the fermionic Casimir effect was considered with the
  more appropriate MIT BC \cite{SantosEtAlJPA2002}.

For simplicity, let us consider a massive fermion field under
anti-periodic BC (in the ${\cal OZ}$ direction) in the presence of a
constant and uniform magnetic field in this same direction. After a
lengthy but straightforward calculation, it can be shown that the
Casimir energy per unit area is given by \cite{Maldito}
\begin{eqnarray}\label{E(a,B)}
{{\cal E}(a,B)\over\ell^2} &=& -\; {2(am)^2\over \pi^2
a^3}\sum_{n=1}^{\infty}
{(-1)^{n-1}\over n^2}\,K_2(amn)  \nonumber \\
&-&{eB\over 4\pi^2 a}\!\!\sum_{n=1}^{\infty}\!(-1)^{n-1}
\!\!\int_0^\infty\!\!\! d\sigma\; e^{-(n/2)^2\sigma-(am)^2/\sigma}\;
\!\!\! L\!\left(\frac{eBa^2}{\sigma}\!\right)\, ,\nonumber
\end{eqnarray}
where we introduced the Langevin function: $L(\xi)= \mbox{coth}\xi
-1/\xi$. In the  strong field limit, we have
\begin{equation}\label{E(B>>)}
  {{\cal E}(a,B)\over \ell^2}\approx -{eBm\over
\pi^2}\sum_{n=1}^{\infty}{(-1)^{n-1}\over n}\,K_1(amn)\; .
\end{equation}
In this limit we can still analyze two distinct situations, namely,
the small mass limit ($ ma\ll 1$) and the large mass limit
 ($ ma\gg 1$). Considering distances between the plates typical
 of Casimir experiments ($a\approx 1\, \mu m)$, we have in the former case
\begin{equation}
{\rho_c(a,B)\over\rho_c(a,0)}\approx 10^{-4}\times{B\over
\mbox{Tesla}}\;\; ;\;\;\;\;ma\ll 1\; ,
\end{equation}
while in the latter case,
\begin{equation}
{\rho_c(a,B)\over\rho_c(a,0)}\approx 10^{-10}\times{B\over
\mbox{Tesla}}\;\; ;\;\;\;\;\;ma\gg 1\; .
\end{equation}
In the above equations $\rho_c(a,B)$ and $\rho_c(a,0)$ are
 the Casimir energy density under  the influence of the
magnetic field and without it, respectively. Observe that for the
Casimir effect of electrons and positrons, which must be treated as
the large mass limit described previously,  huge magnetic fields are
needed in order to enhance the effect (far beyond accessible fields
in the laboratory). In other words, we have shown that though the
Casimir effect of a charged fermionic field can indeed  be altered
by an external magnetic field, the universal constants conspired in
such a way that this influence turns out  to be negligible and
without any chance of a direct measurement at the laboratory.

\pagebreak
\subsection{Magnetic permeability of the constrained Dirac vacuum}

\indent

In contrast to the classical vacuum, the quantum vacuum is far from
being an empty space, inert and insensible to any external
influence. It behaves like a macroscopic medium, in the sense that
it  responds to external agents, as for example electromagnetic
fields or the presence of material plates. As  previously discussed,
recall that the (standard) Casimir effect is nothing but the energy
shift of the vacuum state of the field caused by the presence of
parallel plates. There are many other fascinating  phenomena
associated to the quantum vacuum, namely, the particle creation
produced by the application of an  electric field
\cite{Schwinger1951}, the birefringence of the QED vacuum under an
external magnetic field
\cite{Bialynicki-BirulaPRD1970,AdlerAnnPhys1971} and the Scharnhorst
effect \cite{Scharnhorst1990,Barton1990}, to mention just a few.
This last effect predicts that the velocity of light propagating
perpendicularly to two perfectly conducting parallel plates which
impose (by assumption) BC only on the radiation field is slightly
altered by the presence of the plates. The expected relative
variation in the velocity of light for typical values of possible
experiments is so tiny ($\Delta c/c\approx 10^{-36}$) that this
effect has not been confirmed yet.
 Depending on the nature of the
material plates the velocity of light propagating perpendicular to
the plates is expected to diminish \cite{Cougo-PintoPLB1998}. The
Scharnhorst effect inside a cavity was considered in
\cite{RodriguesPhysicaA2004}.

 The negligible change in the velocity
of light predicted by Scharnhorst may be connected with the fact
that the effect that bears his name is a two-loop QED effect, since
the classical field of a traveling light wave interacts with the
radiation field only through the fermionic loop.

With the purpose of estimating a change in the constitutive
equations of the quantum vacuum at the one-loop level, we were led
to consider the fermionic field submitted to some BC. The magnetic
permeability $\mu$ of the constrained Dirac vacuum was  computed by
the first time in Ref. \cite{Cougo-PintoFarinaTortRafelskiPLB98},
 but with the non-realistic anti-periodic BC.
 In this case, the result found for the relative
 change in the permeability, $\Delta\mu:= \mu - 1$, was also negligible (an
 analogous calculation has also been made in the context of scalar QED
 \cite{Guida2}). However, when the more realistic MIT BC are imposed on the Dirac field
 at two parallel plates things change drastically. In this case, it can be shown
 that the magnetic permeability of the constrained Dirac vacuum is
 given by \cite{BernardinoJPA2006}
$$
\frac{1}{\mu(ma)} = 1 - \frac{\pi - 2}{12 \pi^2}
  {e^2 \over ma} + {1\over 6\pi^2}{e^2 \over ma} H(ma)\, ,
$$
where
\begin{equation}
H(ma)= \int_1^{\infty}dx {x \over (x^2-1)^{3/2}}  \ln \left\{1 +
\left(\frac{x - 1}{x + 1}\right) e^{{- 2 max}} \right\}\, .
\end{equation}
For confining distances of the order of $0,1\mu m$, we have
\begin{equation}
\Delta\mu:=\mu-1\approx 10^{-9}\, .
\end{equation}
The previous value is comparable to the magnetic permeability of
Hydrogen and Nitrogen at room temperature and atmospheric pressure.
Hence, an experimental verification of this result seems to be  not
unfeasible (we will come back to this point in the final remarks).
%
%
%
%
\subsection{The Dynamical Casimir effect}

\indent

As our last topic, we shall briefly discuss  the so called dynamical
Casimir effect, which consists, as the name suggests, of the
consideration of a quantum field in the presence of moving
boundaries. Basically, the coupling between vacuum fluctuations and
a moving boundary  may give rise  to dissipative forces acting on
the boundary as well as to a particle creation phenomenon. In some
sense, these phenomena were expected. Recall that the static Casimir
force is a fluctuating quantity \cite{BartonFluctuatingCasimir} and
hence, using general arguments related to the
fluctuation-dissipation theorem \cite{FlutuacaoDissipacao}
dissipative forces on moving boundaries are expected. Further, using
arguments of energy conservation
 we are led to creation of real particles (photons, if we
are considering the electromagnetic field \cite{Moore70}). For the
above reasons, this topic is sometimes referred to as radiation
reaction force on moving boundaries.

After Schwinger's suggestion that the phenomenon of sonoluminescence
could be explained by the dynamical Casimir effect
\cite{SchwingerLastPapers} (name coined by himself), a lot of work
has been done on this subject. However, it was shown a few years
later that this was not the case (see \cite{MiltonLivro2001} and
references therein for more details).

The dynamical Casimir effect already shows up in the case of one (moving)  mirror
 \cite{FullingDavies76,FordVilenkin82,MaiaNeto94-96}. However, oscillating cavities
  whose walls perform  vibrations in parametric  resonance with a
 certain unperturbed field eigenfrequency may greatly enhance
 the effect \cite{Law94,Dodonov95,Lambrecht96,DodonovKlimov96}.
 Recently, a  one dimensional oscillating
 cavity with walls of different nature was considered
 \cite{AlvesPRA2006}.
 The dynamical Casimir effect
has also been analyzed for a variety of three-dimensional
geometries, including parallel plane plates \cite{Mundarain98},
cylindrical waveguides \cite{waveguide}, and  rectangular
\cite{closed}, cylindrical \cite{cylindrical-cav} and spherical
cavities \cite{spherical}. For a review concerning classical and
quantum phenomena in cavities with moving boundaries see Dodonov
\cite{DodonovReview} and for a variety of topics on
 non-stationary Casimir effect including perspectives of its experimental
 verification  see the special issue \cite{SpecialIssue}.

 In this section, we shall discuss the force exerted by the quantum fluctuations
of a massless scalar field on one moving boundary as well as the
particle creation phenomenon in a unusual example in 1+1 dimensions,
where the field satisfies a Robin BC at the moving boundary
\cite{MintzJPA2006A,MintzJPA2006B}. We shall follow throughout this
paper the perturbative method introduced  by Ford and Vilenkin
\cite{FordVilenkin82}. This method was also applied successfully to
the case of the electromagnetic field under the influence of one
moving (perfectly) conducting plate \cite{MaiaNeto94-96} as well as
an oscillating cavity formed by two parallel (perfectly) conducting
plates \cite{Mundarain98}.

Let us then consider  a massless scalar field $\phi$ in $1+1$
 in the presence of one moving boundary which imposes on the field
 a Robin BC at one moving boundary when observed from a co-moving inertial frame.
 By assumption, the movement of the boundary is  prescribed,
 non-relativistic and of  small amplitude ($\delta q(t)$ is the position
 of the plate at a generic instant $t$). Last assumptions may be
 stated mathematically by
%
$
|\delta \dot{q}(t)|<<c\;\;\;\; \mbox{\color{black} and}
 \;\;\;\;|\delta q(t)|<<c/\omega_0\; ,
 $
%
where $\omega_0$ corresponds to the (main)  mechanical frequency.
Therefore, we must solve the following equation: $\partial^2\phi
(x,t)=0$, with the field satisfying a Robin BC at the moving
boundary given by
\begin{equation}
\left[\frac{\partial}{\partial x} + \delta\dot
q(t)\frac{\partial}{\partial t}\right] \phi(x,t)\vert_{x=\delta
q(t)} = \frac{1}{\beta}\phi(x,t)\vert_{x=\delta q(t)}\; .
\end{equation}
where $\beta$ is a non-negative parameter and the previous condition
was already written in the laboratory frame. We are neglecting terms
of the order ${\cal O}(\delta\dot q^2/c^2)$. The particular cases of
Dirichlet and Neumann BC are reobtained by making $\beta=0$ and
$\beta\rightarrow\infty$, respectively. The dissipative forces for
these particular cases were studied at zero temperature as well as
at finite temperature and also with the field in a coherent state in
\cite{AlvesJPA2003} (dissipative forces on a perfectly conducting
moving plate caused by the vacuum fluctuations of the
electromagnetic field at zero and non-zero temperature were studied
in \cite{PamnPRD2002}). The perturbative approach introduced by Ford
and Vilenkin \cite{FordVilenkin82} consists in writing
\begin{equation}
\phi (x,t)=\phi _{0}(x,t)+\delta \phi (x,t)\; ,  \label{FV}
\end{equation}
where $\phi_0(x,t)$ is the field submitted to a Robin BC at a static
boundary fixed at the origin and $\delta\phi(x,t)$ is the first
order contribution due to the movement of the boundary. The total
force on the moving boundary may be computed with the aid of the
corresponding energy-momentum tensor, namely,
\begin{equation}
\delta F(t)=\langle 0|T^{11}\bigl(t,\delta q^{+}(t)\bigr) -
T^{11}\bigl(t,\delta q^{-}(t)\bigr)|0\rangle \; ,
\label{c_forca_def}
\end{equation}
where superscripts $+$ and $-$ mean that we must compute the
energy-momentum tensor on both sides of the moving boundary. It is
convenient to work with time Fourier transforms. The susceptibility
$\chi(\omega)$ is defined in the Fourier space by
 \begin{equation}
 \delta {\cal F}(\omega )
 =:\chi (\omega )\;\delta Q(\omega )
 \end{equation}
where $ \delta {\cal F}(\omega )$ and $\delta Q(\omega )$ are the
Fourier transformations of $\delta F(t)$ and $\delta  q(t)$,
respectively.
It is illuminating to compute the total work done by the vacuum fluctuations on the moving boundary. It
is straightforward to show that
\begin{equation}
\int_{-\infty}^{+\infty} \delta F(t)\delta \dot q(t)\, dt =
-\frac{1}{\pi}\int_0^\infty d\omega\,\omega\, {\cal
I}m\,\chi(\omega)\vert\delta Q(\omega)\vert^2\; .
\end{equation}
Note that only  the imaginary part of $\chi(\omega)$ appears in the
previous equation. It is responsible for the dissipative effects and
hence it is closely related with the total energy converted into
real particles. On the other hand, the real part of $\chi(\omega)$,
when it exists, does not contribute to the total work and hence it
is not related to particle creation, but to dispersive effects. For
the particular cases of Dirichlet or Neumann BC it can be shown that
the susceptibility is purely imaginary and given by
\begin{equation}
 \chi^D (\omega )=\chi^N (\omega )=i\frac{\hbar \omega ^{3}}{6\pi
c^2}\; ,
 \end{equation}
which implies
 \begin{equation}
  \delta F(t)=\frac{\hbar}{6\pi c^2}\frac{d^3}{dt^3}\delta{q}(t)\, .
\end{equation}
Since ${\cal I}m\,\chi(\omega)>0$, for these cases the vacuum
fluctuations are always dissipating energy from the moving boundary.

\begin{figure}[htbp]
\centerline{\includegraphics[width=7.4cm]{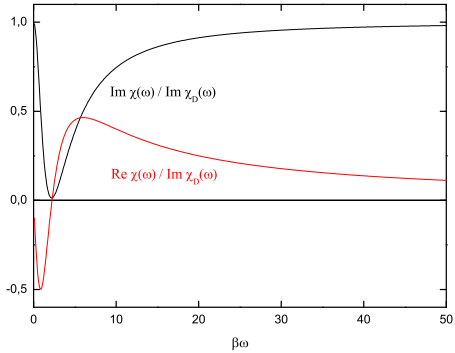}}
\caption{ Imaginary and real parts of $\chi(\omega)$ with Robin BC
appropriately normalized by the value of
 ${\cal I}m\chi(\omega)$ for the Dirichlet BC.}
\label{DispersaoRobin}
\end{figure}

However, for Robin BC an interesting thing happens. It can be shown
that $\chi(\omega)$ acquires also a real part, which gives rise to a
dispersive force acting on the moving boundary. The explicit
expressions of ${\cal R}e\chi(\omega)$ and ${\cal I}m\chi(\omega)$
can be found in \cite{MintzJPA2006A}, but the general behaviour of
them as functions of $\omega$ is shown
 in Figure \ref{DispersaoRobin}.  For convenience, we normalize these quantities
dividing them by the value ${\cal I}m\chi_D(\omega)$, where the
subscript $D$ means that ${\cal I}m\chi(\omega)$ must be computed
with Dirichlet BC.
%
%
%
%
%

Now, let us discuss briefly the particle creation phenomenon under
Robin BC. Here, we shall consider a semi-infinite slab extending
from $-\infty$ to $\delta q(t)$ following as before a prescribed
non-relativistic motion which imposes on the field Robin BC at
$\delta q(t)$. It can be shown that the corresponding spectral
distribution is given by \cite{MintzJPA2006B}
\begin{equation}
\frac{dN}{d\omega}(\omega) = \!\frac{4\omega}
{1+\beta^2\omega^2}\int_0^{\infty}\!\frac{d\omega^{\,\prime}}{2\pi}
\frac{[\delta Q(\omega-\omega^{\,\prime})]^2}
{1+\beta^2{\omega^{\,\prime}}^2} \,\omega^{\,\prime}\!\Biggl[1 \!-
\!\beta^2\omega\omega^{\,\prime}\!\Biggr]^2
\end{equation}
As an explicit example, let us consider the  particular motion
\begin{equation}
\delta q(t)=\delta q_0 \, e^{-\vert t\vert/T}\,\cos(\omega_0 t)\, ,
\end{equation}
where, by assumption, $\omega_0 T\gg 1$ (this is made in order to
single out the effect of a given Fourier component of the motion).
For this case, we obtain the following spectral distribution
\begin{eqnarray}\label{SpectralRobin}
\frac{dN}{d\omega}(\omega)\!\!\!\!\! &=& \!\!(\delta q_0)^2 T
\omega(\omega_0 - \omega)\times\cr
 &\times& \!\!\!\!\!
 \frac{[1-\beta^2\omega (\omega_0 -
\omega)]^2} { (1+\beta^2\omega^2)(1+\beta^2 (\omega_0 -
\omega)^2)}\Theta(\omega_0 - \omega).
\end{eqnarray}
The spectral distributions for the  particular cases of Dirichlet or
Neumann BC can be easily reobtained by making simply $\beta=0$ and
$\beta\rightarrow\infty$, respectively. The results coincide and are
given by \cite{LambrechtPRL96} (we are making $c=1$)
\begin{equation}\label{SpectralRobinDDouNN}
\frac{dN^{(D)}}{d\omega}(\omega) =
 \frac{dN^{(N)}}{d\omega}(\omega) =
 (\delta q_0)^2 T \omega(\omega_0 - \omega)
 \Theta(\omega_0 - \omega)\; .
\end{equation}
A simple inspection in (\ref{SpectralRobin}) shows that,
 due to the  presence of the Heaviside step function, only the
field modes with eigenfrequencies smaller than the mechanical
frequency are excited. Further, the number of particles created per
unit frequency when Robin BC are used is always smaller than the
number of particles created per unit frequency when Dirichlet (or
Neumann) BC are employed.


\begin{figure}[htbp]
\centerline{\includegraphics[width=7.5cm]{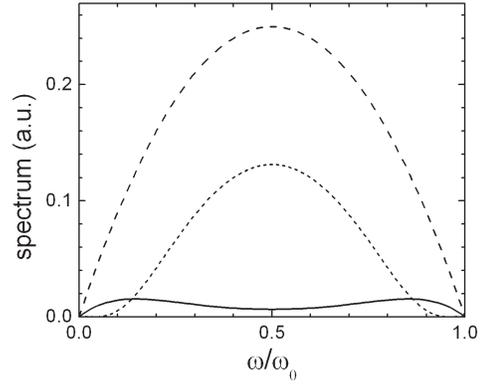}}
 \caption{Spectral distributions of created particles for: Dirichlet and
 Neumann BC (dashed line) and for some interpolating values of
 the parameter $\beta$ (dotted and solid lines).}
\label{Criacao}
\end{figure}


Figure \ref{Criacao} shows the spectral distribution for different
values of the parameter $\beta$, including $\beta=0$ (dashed line),
which corresponds to Dirichlet or Neumann BC. Note that for
interpolating values of $\beta$  particle creation is always smaller
than  for $\beta=0$ (dotted line). Depending on the value of
$\beta$,  particle creation  can be largely suppressed (solid line).

\section{Final Remarks}

\indent

In the last decades there has been a substantial increase in the
study of the Casimir effect and related topics. It is remarkable
that this fascinating effect, considered nowadays as a fundamental
one in QFT, was born in connection with colloidal chemistry, an
essentially experimental science. As we mentioned previously, the
novelty of Casimir's seminal work \cite{Casimir1948} (see also
\cite{Casimir1949}) was the technique employed by him to compute
forces between neutral bodies as is emphasized by Itzykson and Zuber
\cite{ItzyksonZuberLivro}:
\begin{quote}
{\it By considering various types of bodies influencing the vacuum
configuration we may give an interesting interpretation of the
forces acting on them.}
\end{quote}

However, the attractive or repulsive character of the Casimir force
can not be anticipated. It depends on the specific boundary
conditions, the number of space-time dimensions, the nature of the
field (bosonic or fermionic), etc. The \lq\lq mystery{\rq\rq} of the
Casimir effect has intrigued even proeminent physicists such as
Julian Schwinger, as can be seen in his own statement
\cite{MiltonRaadSchwingerAP1978}:
\begin{quote}
 {\it ... one of the least intuitive consequences of quantum
electrodynamics.}
\end{quote}

There is no doubt nowadays about the existence of the (static)
Casimir effect, thanks to the vast list of accurate experiments that
have been made during the last ten years. It is worth emphasizing
that a rigorous comparison between theory and experimental data can
 be achieved only if the effects of temperature and more realistic BC are
 considered. In principle, the former are important in comparison with
 the vacuum contribution for large
 distances, while the latter can not be neglected for short
 distances. Typical ranges investigated in Casimir experiments are form $0.1\mu
m$ to $1.0\mu m$ and, to have an idea of a typical plasma
wavelengths (the plasma wavelength is closely related to the
penetration depth), recall that for  Au we have $\lambda_P\approx
136nm$.

The Casimir effect has become an extremely active area of research
 from both theoretical and experimental points of view and its importance
 lies far beyond the context of QED. This is due to its interdisciplinary
 character, which makes this effect find applications in quantum field theory
 (bag model, for instance), cavity QED, atomic and molecular physics,
 mathematical methods in QFT (development of new regularization and
 renormalization schemes), fixing new constraints in hypothetical forces,
 nanotechnology (nanomachines operated by Casimir forces), condensed
matter physics, gravitation and cosmology, models with compactified
extra-dimensions, etc.

In this work, we considered quantum fields interacting only with
classical boundaries. Besides, these interactions were described by
highly idealized BC. Apart from this kind of interaction, there was
no other interaction present. However, the fields in nature are
interacting fields, like those in QED, etc. Hence, we could ask what
are the first corrections to the Casimir effect when we consider
interacting fields. In principle, they are extremely small. In fact,
for the case of QED, the first radiative correction to the Casimir
energy density (considering  that the conducting plates impose BC
only on the radiation field) was firstly computed by Bordag {\it et
al} \cite{BordagEtAl1985} and is given by
 $ {\cal E}^{(1)}_C(a,\alpha) = {\cal E}^{(0)}_C(a)
\mbox{\Large$\frac{9}{32}\frac{\alpha\lambda_c}{a}$}$,
where $\lambda_c$ is the Compton wavelength of the electron and
${\cal E}^{(0)}_C(a)$ is the zeroth order contribution to the
Casimir energy density. As we see, at least for QED, radiative
corrections to the Casimir effect are experimentally irrelevant.
However, they might be relevant in the bag model, where for quarks
$\lambda_c\approx a$ and also the quark propagators must be
considered submitted to the bag BC \cite{MiltonLivro2001}. Besides,
the study of radiative corrections to the Casimir effect provide a
good laboratory for testing the validity of idealized BC in higher
order of perturbation theory.

Concerning the dynamical Casimir effect, the big challenge is to
conceive an experiment which will be able to detect real photons
created by moving boundaries or by an equivalent physical system
that simulates rapid motion of a boundary. An ingenious proposal of
an experiment has been made recently by the Padova's group
\cite{CasimirDinamicoExp}. There are, of course, many other
interesting aspects of the dynamical Casimir effect that has been
studied, as quantum decoherence \cite{Dechoerence}, mass correction
of the moving mirrors \cite{MassCorrection}, etc. (see also the
reviews \cite{Jaekel-GolestanianReviews}).

As a final comment, we would like to mention that surprising results
have been obtained when a deformed quantum field theory is
considered in connection with the Casimir effect. It seems that the
simultaneous assumptions of deformation and boundary conditions lead
to a new mechanism of creation of real particles even in a static
situation \cite{DeformedQFT-Casimir}. Of course, the Casimir energy
density is also modified by the deformation \cite{Bowes-CougoPinto}.
Quantum field theories with different space-time symmetries, other
than those governed by the usual Poincar\'e algebra (as for example
the $\kappa$-deformed Poincar\'e algebra \cite{Lukierski91}) may
give rise to a modified dispersion relation, a desirable feature in
some tentative models for solving recent astrophysical paradoxes.

\noindent {\bf Acknowledgments:} I am indebted to B. Mintz, P.A.
Maia Neto and R. Rodrigues for a careful reading of the manuscript
and many helpful suggestions. I would like also to thank to all
members of the Casimir group of UFRJ for enlightening discussions
that we have maintained over all these years. Finally, I thank to
CNPq for a partial financial support.

%
%

{

 \end{document}